\documentclass[prd,twocolumn,superscriptaddress,nofootinbib,showpacs]{revtex4}

\usepackage{graphicx,amsmath,amsfonts,amssymb,revsymb,dcolumn,epsfig,bm}

\begin{document}

\title{G\"{o}del-type universes in $\mathbf{f(R)}$ gravity}

\author{M.J. Rebou\c{c}as}\email{reboucas@cbpf.br}
\affiliation{Centro Brasileiro de Pesquisas F\'{\i}sicas,
Rua Dr.\ Xavier Sigaud 150,  \\
22290-180 Rio de Janeiro -- RJ, Brazil}

\author{J. Santos}\email{janilo@dfte.ufrn.br}
\affiliation{Universidade Federal do Rio G. do Norte,
Departamento de F\'{\i}sica,  \\
59072-970 Natal -- RN, Brazil}

\date{\today}

\begin{abstract}
The $f(R)$ gravity theories provide an alternative way to
explain the current cosmic acceleration without
a dark energy matter component. If gravity is governed
by a $f(R)$ theory a number of issues should be reexamined in this
framework, including the violation of causality problem on nonlocal scale.
We examine the question as to whether the $f(R)$ gravity theories
permit space-times in which the causality is violated.
We show that the field equations of these $f(R)$ gravity
theories do not exclude solutions with breakdown of causality for a
physically well-motivated perfect-fluid matter content.
We demonstrate that every perfect-fluid G\"{o}del-type solution of a
generic $f(R)$ gravity satisfying the condition $df/dR > 0$
is necessarily isometric to the G\"odel geometry, and therefore presents
violation of causality. This result extends a theorem on
G\"{o}del-type models, which has been established in the context
of general relativity. We also derive an expression for the critical
radius $r_c$ (beyond which the causality is violated) for an arbitrary
$f(R)$ theory, making apparent that the violation of causality depends
on both the $f(R)$ gravity theory and the matter content.
As an illustration, we concretely take a recent $f(R)$ gravity theory
that is free from singularities of the Ricci scalar and is cosmologically
viable, and show that this theory accommodates noncausal as well as
causal G\"odel-type solutions.
\end{abstract}


\maketitle

\section{Introduction}

The possibility of modifying Einstein's theory of gravitation by adding
terms proportional to powers of the Ricci scalar $R$ to the Einstein-Hilbert
Lagrangian, presently known as $f(R)$ gravity, has a long history (see, e.g.,
\cite{Buchdahl}) and received the attention of many researchers
(see, e.g., Ref.~\cite{Schmidt} for historical reviews).
Quadratic corrections were used to construct a renormalizable gravity
action~\cite{Stelle} and to fuel inflation~\cite{Starobinsky_a}.
Modifications with negative power of $R$ motivated by
string/M-theory~\cite{Nojiri_a}, were also proposed in the scientific
literature. Many of these works were motivated by quantum corrections,
which are important close to the Planck scale.
More recently, due to the impressive amount of astrophysical data
pointing to a phase of accelerated expansion of the Universe~\cite{perl},
$f(R)$ gravity had a revival, motivated by the fact that these theories can
be used to explain the observed accelerating late expansion with no need of
a dark energy component. This has given birth to a great number of
papers~\cite{many_papers} on $f(R)$ gravity (see also Refs.\cite{reviews} for
recent reviews). Several features of these theories,
including solar system tests \cite{Chiba2003}, Newtonian
limit~\cite{Sotiriou2006a}, gravitational stability \cite{Dolgov2003}
and singularities~\cite{Frolov2008}, have been  exhaustively discussed.
General principles such as the so-called energy conditions have also been
used to place constraints on $f(R)$ theory~\cite{energy_conditions}.
As a result, a number of $f(R)$ theories have been suggested
to describe the evolution of the Universe, retaining the standard local
gravity constraints (see, for example, Refs.~\cite{Hu,Starobinsky_b,Ioav}).

If gravitation can be described by a $f(R)$ theory instead of general relativity
(GR), there are a number of issues that ought to be reexamined in the $f(R)$ gravity
framework, including the question as to whether these theories permit space-time
solutions in which the causality is violated.
To tackle this problem in the $f(R)$ gravity framework, we first recall that
there are solutions to the Einstein field equations that possess causal
anomalies in the form of closed time-like curves. The famous solution found by
G\"odel~\cite{Godel49} $60$ years ago is the best known example
of a model that makes it apparent that the general relativity theory
does not exclude the existence of closed timelike
world lines, despite its Lorentzian character which leads to
the local validity of the causality principle.
The G\"odel model is a solution of Einstein's equations
with cosmological constant $\Lambda$ for dust of density
$\rho$, but it can also be interpreted as perfect-fluid
solution (with pressure $p=\rho\,$) without cosmological
constant.
In this regard, it was shown by Bampi and Zordan~\cite{BampiZordan78}
(for a generalization see Ref.~\cite{RebAmanTei86})
that every G\"odel-type solution of Einstein's equations with
a perfect-fluid energy-momentum tensor is necessarily
isometric to the G\"odel spacetime.
Owing to its unexpected properties, G\"odel's model has a
well-recognized importance and has motivated a 
number of investigations on rotating G\"odel-type models
as well as on causal anomalies not only in the context of general
relativity (see, e.g. Refs.~\cite{GT_in_GR})
but also in the framework of other theories of gravitation
(see, for example, Refs.~\cite{GT_Other_Th}).

G\"odel-type universes in gravity theories whose
Lagrangian is an arbitrary function of  the curvature invariants $R$,
$R_{\mu \nu}R^{\mu \nu}$ and $R_{\mu\nu\alpha\beta} R^{\mu\nu\alpha\beta}$
were recently examined by Clifton  and Barrow~\cite{CliftonBarrow2005}.
In particular, they have shown that any $f(R)$ gravity theory in which
$df/dR \neq 0 $, admits a perfect-fluid G\"odel-type solution
with closed timelike curves.%
\footnote{For $f(R)$ gravity with $df/dR=0$,
the existence of these curves depends on the functional
form of $f(R)$, i.e. the violation of causality may or may not
occur~\cite{CliftonBarrow2005}. These theories, however, do
not fulfill the conditions to avoid instabilities and to ensure
agreement with local tests of gravity.}

In this article, to proceed further with the investigation of G\"odel-type
universes along with the question of breakdown of causality in $f(R)$ gravity,
we extend the results of Refs.~\cite{CliftonBarrow2005} and~\cite{Reb_Tiomno}
in four different ways. First,  we examine the dependence of the critical
radius $r_c$ (beyond which the causality is violated) with the $f(R)$ gravity
theory, and derive an expression for the critical radius of G\"{o}del-type
perfect-fluid solutions of any $f(R)$ gravity theory.
Second, we demonstrate that every perfect-fluid G\"{o}del-type solution of a
generic $f(R)$ gravity satisfying the condition%
\footnote{Classically, this condition is necessary to ensure that the effective
Newton constant $G_{eff} = G / f_R$ does not change its sign. At a quantum level,
it prevents the graviton from becoming ghostlike (see, e.g.,
Refs.~\cite{Starobinsky_b} for details).}
$f_R \equiv df/dR > 0$ is necessarily isometric to the G\"odel geometry,
and hence any $f(R)$ gravity
exhibits violation of causality. This extends to the context of $f(R)$
gravity a theorem on G\"{o}del-type models, which has been established in
the framework of general relativity.
Third, given the inevitable breakdown of causality for any perfect-fluid
G\"odel-type solution, we reexamine the violation of causality by considering
two other matter sources, namely combination of a perfect fluid with a scalar
field, and a single scalar field. For both cases we show that $f(R)$ gravity
permit solutions without violation of causality.
Fourth, we concretely illustrate our general results by taking a recent $f(R)$
gravity theory that is free from singularities of the Ricci scalar and is
cosmologically viable~\cite{Ioav}, and show that this theory
accommodates both causal and noncausal solutions.

\section{$\mathbf{f(R)}$ Gravity and G\"{o}del type universes}

The causality problem in $f(R)$ theories can be looked upon as
having two interconnected physically relevant ingredients, namely
the gravity theory (which involves the matter source) and the
space-time geometry.
Regarding the former, we begin by recalling that the action that
defines an $f(R)$ gravity is given by
\begin{equation}  \label{action}
 S=\int d^4x\sqrt{-g}\,\,\left[\,\frac{f(R)}{2\kappa^2} + \mathcal{L}_m
 \right] \,,
\end{equation}
where $\kappa^2\equiv 8\pi G$, $g$ is the determinant of the metric
$g_{\mu\nu}$, $f(R)$ is a function of the Ricci scalar $R$, and
$\mathcal{L}_m$ the Lagrangian density for the matter fields.
Varying this action with respect to the metric we obtain the field equations
\begin{equation}  \label{field_eq}
f_RR_{\mu\nu} - \frac{f}{2}g_{\mu\nu} - \left(\nabla_{\mu}\nabla_{\nu}-
g_{\mu\nu}\,\Box \,\right)f_R = \kappa^2T_{\mu\nu}\,,
\end{equation}
where $f_R\equiv df/dR$, $\Box = g^{\alpha \beta}\,\nabla_{\alpha}\nabla_{\beta}\,$,
$\nabla_{\mu}$ denotes the covariant derivative, and
$T_{\mu\nu} = -(2\,/\!\sqrt{-g}) \,\,\delta (\sqrt{-g}\mathcal{L}_m) / \,
\delta g^{\mu\nu}$ is the matter energy-momentum tensor.
Clearly, for $f(R)=R$ these field equations reduce to the Einstein equations.
An important constraint, often used to simplify the field equations, comes from the
trace of eq.~(\ref{field_eq}), which is given by
\begin{equation}  \label{trace_eq}
3\Box f_R + f_R R - 2f = \kappa^2 T\,,
\end{equation}
where $T\equiv T^{\mu}_{\ \mu}$ is the trace of the energy-momentum tensor.

The second important ingredient of the above mentioned causality problem is
related to the space-time geometry. In this regard, we recall that
the G\"odel-type space-time-homogeneous metrics that we focus our attention on
in this article is given, in cylindrical coordinates [$(r, \phi, z)$],
by~\cite{Reb_Tiomno}
\begin{equation}  \label{G-type_metric}
ds^2 = [dt + H(r)d\phi]^2 - D^2(r)d\phi^2 - dr^2 - dz^2\,,
\end{equation}
where
\begin{eqnarray}
H(r) & = & \frac{4\omega}{m^2}\,\sinh^2(\frac{mr}{2})\,, \label{godel_funH} \\
D(r) & = & \frac{1}{m}\,\sinh(mr)\,, \label{godel_funD}
\end{eqnarray}
with $\omega$ and $m$ being parameters such that $\omega^2 > 0$ and
$-\infty\leq m^2\leq +\infty$.%
\footnote{Clearly, for $m^2 = - \mu^2 <0$ the metric
functions $H(r)$ and $D(r)$ become circular functions
$H(r)=(4\omega/\mu^2)\sin^2(\mu r/2)$ and
$D(r)=\mu^{-1} \sin(\mu r)$, while in the limiting case $m=0$
they become $H= \omega\, r^2$ and $D = r$.}
All G\"odel-type metrics are characterized by the
two parameters $m$ and $\omega$: identical pairs $(m^2, \omega^2)$ specify
isometric space-times~\cite{Reb_Tiomno,TeiRebAman,RebAman}. G\"odel solution
is a particular case of the $m^2 > 0$ class of space-times in which
$m^2= 2 \omega^2$.

The line element of G\"odel-type metrics can also be
written as
\begin{equation} \label{G-type_metric2}
ds^2=dt^2 +2\,H(r)\, dt\,d\phi -dr^2 -G(r)\,d\phi^2 -dz^2 \,,
\end{equation}
where $G(r)= D^2 - H^2$. In this form it is clear that the existence
of closed timelike curves of G\"odel-type,  i.e. circles  defined by
$t, z, r = \text{const}$, depend on the behavior of the function $G(r)$.
If $G(r) < 0$ for a certain range of $r$ ($r_1 < r < r_2$, say)
\emph{G\"odel's circles} defined by  $t, z, r = \text{const}$
are closed timelike curves.
In this regard, it is  easy to show that the causality
features of the  G\"odel-type space-times depend upon
the two independent parameters $m$ and $\omega$~\cite{Reb_Tiomno}.
For $m=0$ there is a critical radius, defined by $\omega r_c = 1$, such
that for all $r>r_c$ there are noncausal G\"odel's circles.
For $m^2= -\mu^2 <0 $  there is an infinite sequence
of alternating causal and noncausal  $t, z, r = \text{const}$ regions
without and with G\"odel's circles. For $0 < m^2 < 4\omega^2$ noncausal
G\"odel's circles occur for $r>r_c$ such that
\begin{equation} \label{r-critical}
\sinh^2 \frac{mr_c}{2}= \left[ \frac{4\omega^2}{m^2} - 1 \right]^{-1}.
\end{equation}
When  $m^2 = 4 \omega^2$ the critical radius $r_c \rightarrow \infty$.
Thus, for $m^2 \geq 4 \omega^2$ there are no G\"odel's circles, and hence
the breakdown of causality of G\"odel-type is avoided.

{}From eqs.~(\ref{G-type_metric}), (\ref{godel_funH}) and (\ref{godel_funD})
it is straightforward to show that the Ricci scalar for the G\"odel-type
metrics takes a constant value $R = 2 (m^2 - \omega^2)$, hence the third
term on the left hand side of equations~(\ref{field_eq}) vanishes.
A further simplification comes about by the following choice of basis:
 \begin{eqnarray}  
\theta^0 &=& dt + H(r)d\phi\,, \quad  \theta^1 = dr\,, \label{one_forms1} \\
\theta^2 &=& D(r)d\phi\,,        \,\quad \qquad \theta^3 = dz \label{one_forms2} \,,
\end{eqnarray}
relative to which the G\"odel-type line
element~(\ref{G-type_metric}) takes the form
\begin{equation}  \label{G-type_metric3}
ds^2 = \eta_{AB}\,\theta^A\theta^B =
(\theta^0)^2 - (\theta^1)^2 - (\theta^2)^2 - (\theta^3)^2\,,
\end{equation}
where clearly $\eta_{AB}=diag(1,-1,-1,-1)$.
Indeed, taking into account the constraint equation~(\ref{trace_eq}), the
field equations (\ref{field_eq}) take the form
\begin{equation}  \label{G_AB-eq}
f_R\,G_{AB} = \kappa^2\,T_{AB} - \frac{1}{2}\,( f + \kappa^2T)\,\eta_{AB}\,,
\end{equation}
where the nonvanishing components of the Einstein tensor $G_{AB}$ take
the quite simple form
\begin{equation} \label{GAB_components}
G_{00} =  3 \omega^2 - m^2, \;\,
G_{11} = G_{22}  =  \omega^2, \;\,
G_{33}  =  m^2 - \omega^2\,.
\end{equation}

Having set up the basic ingredients of the causality problem in $f(R)$
gravity, in the next sections we shall examine whether these theories
permit causal and noncausal solutions.

\section{Noncausal G\"odel-type solution}

An important component of the above gravitational ingredient of the causality
problem is the matter source. In this regard,  we first consider a
physically well-motivated perfect-fluid of density $\rho$ and pressure $p$,
whose energy-momentum tensor in the basis~(\ref{one_forms1})~--~(\ref{one_forms2}) 
is clearly given by
\begin{equation}  \label{perfect_fluid}
T^{(M)}_{AB} = (\rho + p)\,u_A \,  u_B - p\,\, \eta_{AB}\,.
\end{equation}
For this matter source,  the field equations~(\ref{G_AB-eq})
reduce to
\begin{eqnarray}
2(3\omega^2 - m^2)f_R + f &=& \kappa^2\,(\rho + 3p)\,,\label{1st-eq}\\
2\omega^2f_R - f &=& \kappa^2\,(\rho - p)\,, \label{2nd-eq} \\
2(m^2 - \omega^2)f_R - f &=& \kappa^2\,( \rho - p)\,,\label{3rd-eq}
\end{eqnarray}
where we have used eq.~(\ref{GAB_components}).
{}Equations~(\ref{2nd-eq}) and~(\ref{3rd-eq}) give
\begin{equation} \label{f_R-eq}
(2\omega^2 - m^2)f_R = 0\,.
\end{equation}
Thus, for $f(R)$ theories that satisfy the condition
to keep unaltered the sign of the  effective Newton constant
as well as to avoid graviton from becoming ghostlike~\cite{Starobinsky_b},
i.e. $f_R > 0$, equation~(\ref{f_R-eq})
gives $m^2 = 2 \omega^2$, which defines the G\"odel metric, and
the remaining field equations reduce to
\begin{eqnarray}
\kappa^2 p &=& \frac{f}{2} \,,\label{p-eq}\\
\kappa^2 \rho &=& m^2 f_R - \frac{f}{2}  \,, \label{rho-eq}
\end{eqnarray}
where $f$ is an arbitrary function of the Ricci scalar (with $f_R \neq 0$),
and both $f$ and $f_R$ are evaluated at $R=m^2 = 2 \omega^2$.
This result can be seen as an extension of Bampi and Zordan~\cite{BampiZordan78}
result (obtained in the framework of general relativity) to the context of $f(R)$
gravity in the sense that for arbitrary $\rho$ and $p$ (with $p \neq -\rho$)
perfect-fluid  solution of every  $f(R)$ gravity, which satisfies the condition
$f_R>0$, is necessarily isometric to the G\"odel geometry.%
\footnote{We note that this extension is contained in
Ref.~\cite{CliftonBarrow2005} but it has not been explicitly stated.}
Concerning the causality features of these solutions we first note that since
they  are isometric to G\"odel geometry they unavoidably exhibit closed timelike
curves, i.e. noncausal G\"odel's circles whose critical radius $r_c$ is given by
eq.~(\ref{r-critical}). But, taking into account eqs.~(\ref{p-eq}) and~(\ref{rho-eq})
we have that, in the framework of $f(R)$ gravity,  $r_c$ is given by
\begin{equation}  \label{critical_radius}
r_c = \frac{2}{m} \, \sinh^{-1} (1)= 2 \, \sinh^{-1} (1)\,
\sqrt{ \frac{2f_R^{}}{2\kappa^2\rho + f } } \,,
\end{equation}
making apparent that the critical radius, beyond which there exist noncausal
G\"odel's circles, depends on both the gravity theory  and the matter content.
We emphasize this expression~(\ref{critical_radius}) for the critical radius
holds for any $f(R)$ gravity which satisfies the condition $f_R > 0$.

Despite this inescapable breakdown of causality for any perfect-fluid 
G\"odel-type $f(R)$ solution, to concretely illustrate an estimation of  the bounds
on $r_c$ for a specific theory, let us consider the recently proposed $f(R)$ theory
described by~\cite{Ioav}
\begin{equation}  \label{Ioav-eq}
f(R) = R - \alpha R_{\ast}\ln{\left( 1 + \frac{R}{R_{\ast}} \right)}\,,
\end{equation}
which is free from singularities of the Ricci scalar, cosmologically viable
and satisfies the existence of relativistic stars for
positive parameters $\alpha$ and $R_{\ast}$.
To this end, we use the positivity of the density  $\rho$ and
eq.~(\ref{rho-eq}) to obtain
\begin{equation} \label{posrho}
m^2 f_R - \frac{f}{2} \geq 0 \,,
\end{equation}
where $f$ is an arbitrary function of the $R$ (with $f_R \neq 0$), and
both $f$ and $f_R$ are evaluated at $R=m^2$. By using~(\ref{Ioav-eq}) for
$\alpha =2 $ (see Ref.~\cite{Ioav}) it is easy to show that the
inequality (\ref{posrho}) holds for all $m$ such that
$m^2 \geq 0.55 R_{\ast}$, making therefore explicit the lower bound on
$m^2$ and therefore on the critical radius $r_c$ for this theory.

\section{Causal G\"odel-type solution}

Since any perfect-fluid G\"odel-type solution of $f(R)$ gravity is
inevitably noncausal, the question as to whether other matter
sources could generate G\"odel-type causal solutions naturally arises
at this point. In this section we shall examine this problem by
considering two different matter sources, namely a combination of
a perfect fluid with a scalar field, and a single scalar field.%
\footnote{We note that the presence of a single closed timelike curve
as, for example, a G\"odel's circle, is an unequivocal manifestation of
violation of causality. However, a space-time may admit noncausal
curves other than G\"odel's circles. Therefore, throughout this paper
by causal solutions we mean solutions with no violation of
causality of G\"odel-type, i.e., no G\"odel's circles.}

\subsection{Perfect fluid plus Scalar field}

The combined energy-momentum tensor we consider is given by
\begin{equation} \label{TAB-combined}
T_{AB} = T^{(M)}_{AB} + T^{(S)}_{AB} \,,
\end{equation}
where $T^{(M)}_{AB}$  corresponds to a  perfect fluid
[eq.~(\ref{perfect_fluid})] and $T^{(S)}_{AB}$ is
energy-momentum tensor of a scalar field, i.e.
\begin{equation}
T^{(S)}_{AB}= \Phi^{}_{|A}\,\Phi^{}_{|B} - \frac{1}{2}\,\eta^{}_{AB}\,
\Phi^{}_{M} \,\Phi^{}_{N}\, \eta^{MN} ,
\end{equation}
where a vertical bar denotes components of covariant derivatives relative to
the local basis $\theta^A = e^{(A)}_\alpha \, dx^\alpha $ 
[see eq.~(\ref{one_forms1}) and~(\ref{one_forms2}) ].
Following Ref.~\cite{Reb_Tiomno} it is straightforward to show that a scalar field
of the form $\Phi (z)= e z + \text{const}$ satisfies the scalar field equation
$\Box \,\Phi = \eta^{AB}_{}\,\nabla_{A} \nabla_{B} \,\Phi\,=0$
for a constant amplitude $e$ of $\Phi (z)$. Thus, the nonvanishing
components of energy-moment tensor for this scalar field are
\begin{equation}  \label{S-comp}
T^{(S)}_{00} = - T^{(S)}_{11} = - T^{(S)}_{22} = T^{(S)}_{33} = \frac{e^2}{2}\,,
\end{equation}
and the field equations (\ref{G_AB-eq}) for the combined matter source
(\ref{TAB-combined}) can be written in the form
\begin{eqnarray}
\kappa^2 e^2 & = & (m^2 - 2\omega^2)f_R\,,\label{1st-eq-gen} \\
\kappa^2\,p & = & \frac{1}{2}\,(2\omega^2 - m^2)f_R + \frac{f}{2}\,,\label{2nd-eq-gen}\\
\kappa^2\rho & = &  \frac{1}{2}\,(6\omega^2 - m^2)f_R - \frac{f}{2}\,,\label{3rd-eq-gen}
\end{eqnarray}
where $f$ is an arbitrary function of the Ricci scalar (with $f_R \neq 0$),
and both $f$ and $f_R$ are evaluated at $R= 2(m^2 - \omega^2)$.
A causal G\"odel-type class of solutions of these equations that
satisfies the condition $f_R> 0$  is given by
\begin{eqnarray}
m^2 &=& 4 \omega^2 \label{1st-comb}\,, \\
f_R  & = & \frac{\kappa^2 e^2}{2 \,\omega^2 } \,,\label{2nd-comb} \\
\kappa^2\,p &=& - \kappa^2\rho= -\omega^2 f_R + \frac{f}{2}\,,\label{3rd-comb}
\end{eqnarray}
where from equations (\ref{r-critical}) and (\ref{1st-comb}) one
clearly has that the critical radius $r_c \rightarrow \infty$.
Hence, for this combination of matter fields, there is no violation of causality
of  G\"odel type (G\"odel's circles) for any $f(R)$ gravity that satisfies
the conditions $f_R> 0$.

As an illustration, we shall now concretely examine whether the
theory described by~(\ref{Ioav-eq}) admits this type of causal solution.
For this theory, eq.~(\ref{2nd-comb}) gives rise to a quadratic equation in
the variable $m^2/R_{\ast}$ whose roots are given in terms of $e^2/R_{\ast}$
by
\begin{equation}  \label{roots}
\frac{m^2_{\pm}}{R_{\ast}} = \frac{1}{3}\,\left[ 1+3\frac{\kappa^2e^2}{R_{\ast}} \pm \sqrt{1+18\frac{\kappa^2e^2}{R_{\ast}}
+ 9\left(\frac{\kappa^2e^2}{R_{\ast}} \right)^2} \right]\,,
\end{equation}
where we have taken $\alpha =2$ (see Ref.~\cite{Ioav} for details).
Clearly, the  positivity of the density parameter $\rho$ [as given by~(\ref{3rd-comb})
for $f$ evaluated at $R=6\omega^2 = 3\,m^2/2$ and $f_R$ given by (\ref{2nd-comb})]
is assured  by $\kappa^2 e^2 - f \geq 0$ for each root of equation~(\ref{roots}).
Regarding the first root $m^2_{+}/R_{\ast}$ the positivity of $\rho$ gives
\begin{equation}  \label{positive-root}
0\leq \frac{\kappa^2e^2}{R_{\ast}}\lesssim 0.8 \quad \mbox{and} \qquad
0.7\lesssim \frac{m^2}{R_{\ast}}\lesssim 2.7\,.
\end{equation}
Thus, for values $\kappa^2e^2/R_{\ast}$ and $m^2/R_{\ast}$ within these intervals
there are causal solutions of the $f(R)$ gravity of Ref.~\cite{Ioav}
generated by the combination of a perfect fluid with a scalar field
such that $\rho \geq 0$.%
\footnote{For completeness, we mention that the second root of~(\ref{roots}),
i.e. $m^2_{-}/R_{\ast}$ along with the positivity of $\rho$ furnishes
$ \kappa^2 e^2 / R_{\ast} \gtrsim 2.5 $ and $m^2 / R_{\ast}< 0$.
Negative values of $m^2$ are known to lead to violation of causality with
alternating causal and noncausal G\"odel's circles~\cite{Reb_Tiomno,Novello}.}

\subsection{Scalar field}

For the scalar field $\Phi(z)$ as the single source component, and $f_R \neq 0$,
the field equations~(\ref{1st-eq-gen})~--~(\ref{3rd-eq-gen}) give rise to
the unique class of G\"odel-type solutions
\begin{eqnarray}
m^2 &=& 4 \omega^2 \label{1st-phi-a}\,, \\
f_R  & = & \frac{\kappa^2 e^2}{2 \,\omega^2 } \,,\label{2nd-phi-b} \\
f &=&  k^2 e^2 \,,\label{3rd-phi-c}
\end{eqnarray}
where $f$ is an arbitrary function of $R$ (with $f_R \neq 0$), and both
$f$ and $f_R$ are evaluated at $R= 2(m^2 - \omega^2)$.
This clearly defines a class of solutions with no violation of
causality of G\"odel type ($r_c \rightarrow \infty$) for an arbitrary
$f(R)$ with $f_R \neq 0$.

As an illustration, we note that for this source, the $f(R)$ theory
described by~(\ref{Ioav-eq}) also permits a causal solution. Indeed, as
eq.~(\ref{2nd-phi-b}) is identical to eq.~(\ref{2nd-comb}) it clearly
has two roots given by eq.~(\ref{roots}).
Inserting the first root, $m^2_{+}/R_{\ast}$, into~(\ref{3rd-phi-c})
one finds the following values:
\begin{equation}  \label{SF-root-posit}
\frac{\kappa^2e^2}{R_{\ast}}\thickapprox 0.82 \qquad \mbox{and} \qquad
\frac{m^2}{R_{\ast}} \thickapprox 2.7\,,
\end{equation}
making apparent that the theory of Ref.~\cite{Ioav} accommodates
the solution given by (\ref{1st-phi-a})~--~(\ref{SF-root-posit}), which
has no violation of causality of G\"odel type.%
\footnote{The  second root of~(\ref{roots}) gives
$\kappa^2e^2 / R_{\ast} \thickapprox 2.44$ and
$m^2 / R_{\ast} \thickapprox -0.53$, which leads again to violation of
causality with alternating causal and noncausal circles~\cite{Novello,Reb_Tiomno}.}

\section{Concluding Remarks}

The so-called $f(R)$ gravity theory provides an alternative way to
explain the current cosmic acceleration with no need of invoking
either the existence of an extra spatial dimension or a dark energy
component.
If gravity is governed by a $f(R)$ theory instead of general relativity,
various issues should be reexamined in the $f(R)$ framework. This includes
the breakdown of causality.
In $f(R)$ gravity theories the causal structure of four-dimensional
space-time has locally the same qualitative nature as the
flat space-time of special relativity --- causality holds locally.
The nonlocal question, however, is left open, and violation of
causality can occur.

In this article, we have examined the question as to whether the $f(R)$
gravity theories permit  space-times in which the causality
is violated or not, and generalize the results of
Refs.~\cite{CliftonBarrow2005} and ~\cite{Reb_Tiomno}.
For physically well-motivated perfect-fluid matter sources, we showed
that every perfect fluid (with $p \neq \rho$) G\"{o}del-type solution of
an arbitrary $f(R)$ gravity that satisfies the the condition $f_R>0$ is
necessarily isometric to the G\"odel geometry, making explicit
that the violation of causality is unavoidable feature of
any $f(R)$ gravity. This results is a generalization
of the Bampi-Zordan theorem~\cite{BampiZordan78}  which has
been established in the context of Einstein's theory of gravitation.
We have derived an expression for the critical
radius $r_c$ (beyond which the causality is violated) for an arbitrary
$f(R)$ theory (with $f_R \neq 0$), making apparent that the functional
character of the violation of causality depends on both the $f(R)$
gravity theory and the matter content.

We have also examined the question as to whether other matter
sources could give rise to G\"odel-type causal solutions by considering
a combination of perfect fluid with a scalar field, and simply a single
scalar field. We have shown that in both cases  G\"odel-type solutions of
an arbitrary $f(R)$ theory (with $f_R \neq 0$) with no violation of causality
are permitted. We have also found a general class of such causal solution
for an arbitrary $f(R)$ theory that satisfies the condition $f_R>0$.
As an illustration, we  have concretely considered a recent $f(R)$ gravity theory
that is free from singularities of the Ricci scalar and is cosmologically
viable~\cite{Ioav}, and showed that this theory accommodates both noncausal and
causal G\"odel-type solutions.

\vspace{-0.5cm}
\begin{acknowledgments}
This work is supported by Conselho Nacional de Desenvolvimento Cient\'{i}fico e
Tecnol\'ogico (CNPq) -- Brasil, under Grant No.\ 472436/2007-4.
M.J.R. and J.S. thank CNPq for the research grants under which this work
was carried out.
J.S. is supported by PRONEX (CNPq/FAPERN) and also thanks CNPq support
under the Grant No.\ 479469/2008-3.
We are grateful to A.F.F. Teixeira for indicating omissions and misprints.
We also thank S.E. Perez Bergliaffa for fruitful discussions.
\end{acknowledgments}


\end{document}